\documentclass{statsoc}
\usepackage{geometry}
\usepackage{vmargin} 

\usepackage{mathtools}
\usepackage{booktabs}
\usepackage{amsmath,amsfonts,amssymb}
\usepackage{natbib}
\usepackage{graphics}
\usepackage{subfigure}
\usepackage{chngcntr}
\usepackage[english]{babel} 

\usepackage{rotating} 
\usepackage{afterpage}
\usepackage{footnote}
\usepackage[toc,page]{appendix}
\usepackage{color}
\usepackage{floatrow}
\usepackage{subfloat}
\usepackage{caption}
\newfloatcommand{capbtabbox}{table}[][\FBwidth]

\makeatletter
\newcommand{\bi}{\begin{itemize} \itemsep0em}
\newcommand{\ei}{\end{itemize}}
\newcommand{\distas}[1]{\mathbin{\overset{#1}{\kern\z@\sim}}}%
\newcommand{\bm}[1]{\mathbf{#1}}
\newsavebox{\mybox}\newsavebox{\mysim}
\addto{\captionsenglish}{\renewcommand{\abstractname}{Summary}}

\makeatother

\let\oldbibliography\thebibliography
\renewcommand{\thebibliography}[1]{\oldbibliography{#1}
\setlength{\itemsep}{0pt}} 

\title{A regional compound Poisson process for hurricane and tropical storm damage} 

\author{Simon Mak\footnote{Corresponding author}}
\address{Georgia Institute of Technology, Atlanta, USA.}
\email{smak6@gatech.edu}
\author{Derek Bingham}
\address{Simon Fraser University, Burnaby, Canada.}
\email{dbingham@stat.sfu.ca}
\author[Mak {\it et al.}]{Yi Lu}
\address{Simon Fraser University, Burnaby, Canada.}
\email{yilu@stat.sfu.ca}

\begin{document}

\maketitle 


\begin{abstract}
In light of intense hurricane activity along the U.S. Atlantic coast, attention has turned to understanding both the economic impact and behaviour of these storms. The compound Poisson-lognormal process has been proposed as a model for aggregate storm damage, but does not shed light on regional analysis since storm path data are not used. In this paper, we propose a fully Bayesian regional prediction model which uses conditional autoregressive (CAR) models to account for both storm paths and spatial patterns for storm damage. When fitted to historical data, the analysis from our model both confirms previous findings and reveals new insights on regional storm tendencies. Posterior predictive samples can also be used for pricing regional insurance premiums, which we illustrate using three different risk measures.
\end{abstract}


\section{Introduction} 

In wake of the devastation caused by intense hurricane and tropical storm (HTS) activity, attention has turned to understanding the tendencies and economic impact from these storms. In particular, insurance companies have expressed interest in developing damage prediction models for these catastrophes (e.g., \citealp{Cha1996}; \citealp{Mus1997}), since such models allow for risk assessment and premium pricing of insurance products. However, to price insurance premiums for each localized, or regional, location (hereafter ``regional premiums''), models that predict only damage aggregated over all locations (e.g., \citealp{KVG1988,PSW2006}) are of little use, since they do not account for regional variation. We view these two contrasting goals (i.e., prediction of aggregate vs. localized damage) as macro-level and micro-level, respectively. In this paper, a new micro-level model is proposed that incorporates storm path data and, to the best of our knowledge, is the first to provide regional HTS damage forecasts (where regions here are affected U.S. states).\\

One of the first HTS damage models proposed was the compound Poisson-lognormal model \citep{Kat2002}. For this model, the total damage incurred up to time $t$, $X(t)$, is modelled as:
\begin{equation} X(t) = \sum_{i = 1}^{N(t)} Y^{(i)}, \label{eq:CPL} \end{equation}
where $\{N(t), t \geq 0\}$ is a Poisson process for storm counts, and $\{Y^{(i)}\}_{i = 1}^{\infty}$ are i.i.d. lognormal-distributed monetary damage incurred for each storm, independent of $\{N(t), t \geq 0\}$. This specification is a common model for aggregate loss in certain lines of insurance (e.g., \citealp{ZDL2006}), as well as banking operational risk models \citep{LSD2009} and total rainfall models \citep{Tho1984}. However, since our goal is to develop a micro-level model that incorporates all historical storm data available by location, we cannot use the framework in \eqref{eq:CPL} directly, since it neglects the underlying path from each storm.\\

El Ni\~no - Southern Oscillation (ENSO) phases, which categorize sea surface temperatures (SST) in the tropic Pacific Ocean, can also have a notable impact on HTS behaviour (e.g., \citealp{PL1998, Cea2007, Pie2009}), and should therefore be included as covariates for prediction. \cite{Cea2011} explored models with various ENSO and seasonal effects for both storm count and damage, and found that, while ENSO and seasonality have a noticeable impact on counts, only ENSO effects appear to impact damage. Unfortunately, since their model is an extension of \eqref{eq:CPL} with covariates, it also cannot serve as a micro-level model. Their model also examines only storm losses above a certain threshold, whereas our interests lie in modeling for the entire loss distribution.\\

In this paper, we aim to incorporate all available HTS data - storm paths (defined as the set of regional locations hit by a storm), ENSO phase information and regional damage effects, to jointly predict HTS counts and regional damage. For spatial analysis, the proposed model incorporates conditional autoregressive (CAR) models (\citealp{Bes1974}), along with \textit{line-of-sight (LOS) connectivity}, which allows storm path information to be borrowed between locations that do not share a common land boundary (i.e., not physically connected). This new notion of connectivity is motivated by the fact that all historical storm paths impact states that are LOS connected, but some are not physically connected. LOS connectivity is therefore crucial for providing realistic storm path predictions. When fit to historical HTS data \citep{Pea2008}, the proposed model confirms several previous findings, including the strong ENSO impact on storm damage (e.g., \citealp{PL1998,Kat2002}) and on seasonal patterns for storm counts (\citealp{LZ2012}). More importantly, by integrating storm path data, our model also reveals new insights on regional storm behaviour, including ENSO effects on the spatial patterns for both storm path and damage. Regional damage predictions from the proposed model can be used not only by insurers to provide local risk assessment and to price regional premiums, but also by government policymakers to guide decisions on infrastructure and disaster prevention.\\

The paper is organized as follows. Section 2 gives a brief exploratory analysis of historical HTS data in \cite{Pea2008}. Section 3 provides an overview of Gaussian and autologistic CAR models. Section 4 presents the three components (storm count, path and damage) of the proposed model, and discusses prior specifications and posterior exploration for each component. Section 5 outlines a simulation study, and Section 6 discusses our findings after fitting the proposed model to historical data. We conclude the paper with some future research directions in Section 7.


\section{Exploratory analysis of HTS data}

\begin{figure}[!t]
\centering
\includegraphics[scale = 0.48]{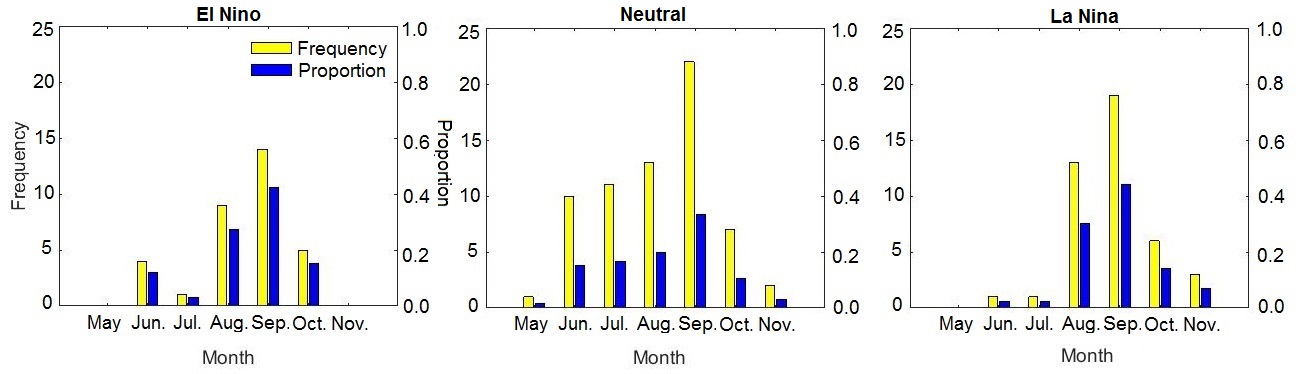}
\caption{\label{fig:seasoninteract}Bar-charts of monthly HTS frequency and proportions for each ENSO phase.}
\end{figure}

\subsection{HTS data and ENSO phases}
Data on occurrence times, storm paths and normalized economic losses of HTS along the U.S. Gulf and Atlantic coast from 1900-2005 are reported in \cite{Pea2008}. Although the set of regional locations provided in this dataset is the set of U.S. states, we refer to these as \textit{locations} from here on to provide a more general exposition of the model. Economic loss in this report was defined as ``direct losses associated with a hurricane's impact as determined in the weeks after the event'' (this excludes indirect losses such as demand surges, loss mitigation or other longer macroeconomic effects), and we adopt the same definition here. To provide meaningful comparisons between storms occurring at different times, original reported losses were adjusted to 2005 conditions by a normalization method (called PL05) that accounts for changes in inflation, real-wealth-per-capita and population. We use these normalized losses to fit the proposed model, implying that damage forecasts are for 2005 conditions as well.\\

ENSO phases classify SST variation in the tropic Pacific into three categories: El Ni\~no (warm anomalies), Neutral (no anomalies) and La Ni\~na (cold anomalies). In particular, the National Oceanic and Atmospheric Administration (NOAA) defines a year to be El Ni\~no, Neutral or La Ni\~na if the three-month temperature average in August, September and October (ASO) is above, within or below $0.5 \,^{\circ}\mathrm{C}$ normal for that period  (e.g., \citealp{PL1998}). Although the official hurricane season runs from Jun. 1st to Nov. 30th, only the three months ASO are used in this definition, since over 95\% of high-intensity HTS occur within this timeframe \citep{Lan1993}. Unfortunately, NOAA's categorization is only available from 1950 onwards, and replication of their classification for prior years is difficult due to a changing base-period methodology. {In order to incorporate reliable ENSO phase data, we are therefore restricted to analyzing storms occurring after 1950. The final dataset consists of 142 storms from 1950-2005: 33 in El Ni\~no years, 66 in Neutral and 43 in La Ni\~na.}


\subsection{Exploratory data analysis}
\floatsetup[figure]{style=Boxed,frameset={\fboxrule1pt}}
\begin{subfigures}
\begin{figure}[t]
\begin{floatrow}
\ffigbox{\includegraphics[scale = 0.60]{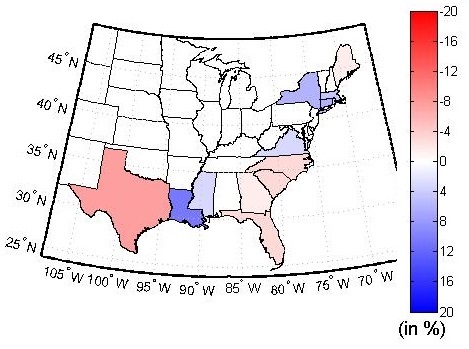}}
{\caption{Difference between El Ni\~no and Neutral hit-rates by location. Blue indicates a higher El Ni\~no rate, and red indicates a higher Neutral one.}
\label{fig:histpath}
}
\hspace{0.2cm}
\ffigbox{\includegraphics[scale = 0.75]{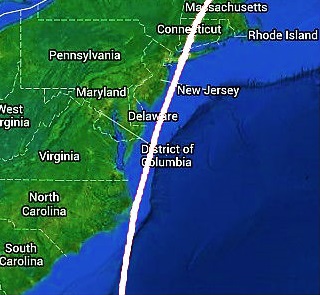}}
{\caption{The storm path for Hurricane Gloria (NC, NY, CT, MA) is not spatially connected, but is LOS connected.}
\label{fig:gloria}
}
\end{floatrow}
\end{figure}
\end{subfigures}
\floatsetup[figure]{style=plain}

Here, we make several observations about HTS counts, paths and damage which motivate the proposed model. {The average storm count in El Ni\~no, Neutral and La Ni\~na years is 2.06, 2.44 and 3.31, with standard deviations of 1.61, 1.55 and 1.80, respectively, which suggests that colder SST anomalies tend to increase both storm activity and variation}. From Figure \ref{fig:seasoninteract}, which plots bar-charts of HTS frequency and proportion for each month, a majority of La Ni\~na and El Ni\~no storms can be seen to occur in ASO, whereas in Neutral years, storms are more evenly spaced over the hurricane season. This gives visual evidence for differing seasonal patterns in each ENSO phase, which we call \textit{ENSO-seasonal count interactions}. \cite{LZ2012} also note a similar interaction effect using a three-state hidden Markov model (HMM) for count intensity, with fitted intensities (Figure 6 in their paper) closely resembling continuous analogues of Figure \ref{fig:seasoninteract}. This suggests that the three ENSO phases may indeed be the HMM states, and highlights the need to model ENSO-seasonal interactions.\\

Next, looking at storm paths in the eastern United States, it turns out that certain locations, e.g., Florida (FL) and Texas (TX), are more often hit than others, e.g., Maine (ME) and Rhode Island (RI), for all three ENSO phases (see Figure \ref{fig:legend} in the Appendix for state map and abbreviations). {Indeed, Moran's $I$ test \citep{Mor1950} returns p-values of 0.07, 0.01 and 0.31 against the null hypothesis of no spatial autocorrelation for storm hit-rates in El Ni\~no, Neutral and La Ni\~na, respectively. This provides, at least for the first two phases, some evidence for spatial hit-rate dependencies, and suggests a spatial component is needed for modeling storm paths. Moreover, from Figure \ref{fig:histpath}, which plots the difference between historical hit-rates in El Ni\~no and Neutral years in each location (with blue indicating higher El Ni\~no hit-rates, and red indicating higher Neutral rates), a subtle ENSO effect is present on spatial patterns as well. Storms in Neutral years concentrate on locations along the southern coast, whereas storms in El Ni\~no tend to move towards northern locations and Louisiana.} We call these effects \textit{ENSO-spatial path interactions}.\\

Several historical storms also hit locations that are not physically connected, i.e., there are states on these paths which do not share a common land boundary with any other locations along the path. For example, the path for Hurricane Gloria, plotted in Figure \ref{fig:gloria}, hits states that are not physically connected insofar since NC does not share a land boundary with the other three states on the hurricane's path. Connectedness is an important feature in path prediction, because we do not want to predict storm paths which cannot occur in reality. Consequently, we introduce the key idea of \textit{line-of-sight (LOS) connectedness}: two locations are \textit{LOS neighbours} if there exists an unobstructed sight line from one to the other, and a path is \textit{LOS connected} if it is connected on the graph induced by LOS neighbours. Under this definition, all historical storm paths are LOS connected. This new notion of connectivity is particularly important for the proposed model, since it not only allows for more realistic path predictions, but also permits regional path information to be shared between more locations.\\

Lastly, for storm damage (measured in log-U.S. \$), La Ni\~na storms incurred the highest median log-damage per storm (20.3), followed by Neutral (19.6) and El Ni\~no (18.8), which is in line with previous findings on the impact of ENSO on storm damage (e.g., \citealp{PL1998}). Spatial patterns for damage are also of interest in our model, since these allow for regional analysis and prediction. Table \ref{tbl:EDAdamage2} provides the median and standard deviation of log-damage for the 122 single-location storms, split by location and ENSO (multiple-location storms are not included, since the data only provides aggregated damage over all locations). Although there is some evidence for spatial effects, e.g., FL incurring higher median damage (19.3) than LA (19.1) or TX (19.0), it is difficult to judge if these effects are genuine or due to random error, since many locations contain little to no storm data. We are also interested in \textit{ENSO-spatial damage interactions}, which account for varying spatial patterns of damage for each ENSO phase, but this is even harder to detect from a casual inspection of Table \ref{tbl:EDAdamage2}, since nearly half of ENSO-location combinations have no data. To estimate these effects, a model incorporating spatial information is needed to borrow data from neighbouring locations.

\begin{table}
\ttabbox{
\centering
\small
\begin{tabular}{ c  c  c  c  c  c  c  c  c  c  c  c  c  c  c }
\toprule
& \multicolumn{1}{c}{ \textit{ME}} & \multicolumn{1}{c}{ \textit{MA}} & \multicolumn{1}{c}{ \textit{RI}} & \multicolumn{1}{c}{ \textit{CT}} & \multicolumn{1}{c}{ \textit{NY}} & \multicolumn{1}{c}{ \textit{VA}} & \multicolumn{1}{c}{ \textit{NC}} \\ 
\toprule
El Ni{\~n}o (25) 	& - 			& 17.5,NA(1) 	& - & - & 20.0,NA(1)& - 		& 15.3,3.0(3) \\ 
Neutral (58) 		& 16.5,NA(1) 	& 19.4,NA(1) 	& - & - & - 		& - 		& 20.0,1.6(8) \\ 
La Ni{\~n}a (39) 	& - 			& 21.8,NA(1) 	& - & - & - 		& - 		& 21.3,1.6(8)\\ 
\hline
\textbf{Total (122)} & 16.5,NA(1) 	& 19.4,2.2(3) 	& - & - & 20.0,NA(1)& - 		& 20.1,1.2(19) \\ 
\toprule
\multicolumn{1}{c}{}\\
& \multicolumn{1}{c}{ \textit{SC}} & \multicolumn{1}{c}{ \textit{GA}} & \multicolumn{1}{c}{ \textit{FL}} & \multicolumn{1}{c}{ \textit{AL}} & \multicolumn{1}{c}{ \textit{MS}} & \multicolumn{1}{c}{ \textit{LA}} & \multicolumn{1}{c}{ \textit{TX}}\\
\toprule
El Ni{\~n}o (25) 	& 17.2,2.2(2) & - & 21.2,3.6(8)   & - & 18.6,2.2(2) 	& 19.0,1.4(4) & 16.1,1.8(4)\\
Neutral (58)  	& 19.0,2.1(5) & - & 19.1,1.8(18) & - & 17.2,NA(1) 	& 19.2,1.8(8) & 19.3,2.1(16)\\
La Ni{\~n}a (39)  	& - 		 & - & 20.3,2.6(17) & - & - 		& 17.5,3.0(4) & 18.8,2.4(9)\\
\hline
\textbf{Total (122)}  & 18.8,2.3(7) & - & 19.3,2.5(43) & - & 17.2,1.8(3) & 19.1,2.0(16) & 19.0,2.4(29)\\
\toprule
\end{tabular}
}
{\caption{Median and standard deviation of log-damage for single-location storms at each location and ENSO phase. Number of storms is bracketed.}
\label{tbl:EDAdamage2}
}
\end{table}


\section{Notation and background}

To foreshadow the storm path and damage models, we provide a brief overview of Gaussian and autologistic CAR models; a more detailed discussion can be found in \cite{BCG2004}. Let $V = \{1, \cdots, S\}$ index a set of $S$ areal locations, $E$ be a set of edges indicating spatial connectivity of locations in $V$, and $G = (V,E)$ be the resulting (undirected) graph. Also, let $\boldsymbol{\gamma}_{\cdot,s} = (\gamma_{1,s}, \gamma_{2,s}, \cdots, \gamma_{I,s})^T$ be a vector of $I$ measurements at location $s$, and let $\boldsymbol{\gamma} = (\boldsymbol{\gamma}_{\cdot,1}^T, \cdots, \boldsymbol{\gamma}_{\cdot,S}^T)^T$ be the vector of measurements over all $S$ locations. A \textit{multivariate Gaussian CAR model}, denoted as ${MCAR(\mathbf{\Sigma})}$, is given by the conditional specification
\begin{equation}
\label{eq:MCARfullcond}
\boldsymbol{\gamma}_{\cdot,s} \; | \; \boldsymbol{\gamma}_{\cdot,(-s)}, \bm{\Sigma} \sim N\left( \frac{1}{n_s}\sum_{r \sim s} \boldsymbol{\gamma}_{\cdot,r},\frac{1}{n_s} \bm{\Sigma} \right), \qquad \forall s = 1, \cdots, S,
\end{equation}
where $\boldsymbol{\gamma}_{\cdot,(-s)}\equiv \{\boldsymbol{\gamma}_{\cdot,r} : r \neq s\}$ is the set of measurements excluding those at location $s$, $\mathbf{\bm{\Sigma}}$ is the unscaled conditional covariance matrix of $\boldsymbol{\gamma}_{\cdot,s}$, and $n_s$ is the number of neighbours at location $s$. The operator $\sim$ indicates adjacency on $G$, i.e., $r \sim s$ if and only if the edge $(r,s)$ is in $E$. By Brook's Lemma \citep{Bro1964}, the joint density of \eqref{eq:MCARfullcond} becomes
\begin{align}
\label{eq:MCARjoint2}
\begin{split}
p(\boldsymbol{\gamma}) \propto \textrm{exp} \left\{- \frac{1}{2} \boldsymbol{\gamma}^T [(\mathbf{D} - \mathbf{W}) \otimes \bm{\Sigma}^{-1}] \boldsymbol{\gamma} \right\},
\end{split}
\end{align}
where $\otimes$ is the Kronecker product operator, $\mathbf{W}$ is the adjacency matrix of $G$, and $\mathbf{D} = \textrm{diag}\left\{n_s\right\}$. That is, under the conditional specification \eqref{eq:MCARfullcond}, $\boldsymbol{\gamma}$ follows a Gaussian distribution with inverse covariance $(\mathbf{D} - \mathbf{W}) \otimes \bm{\Sigma}^{-1}$. \\

MCAR models make for convenient spatial priors in Bayesian models for two reasons. First, the specification \eqref{eq:MCARfullcond} often allows for direct sampling in each Gibbs sampling step (Chapter 3 in \citealp{BCG2004}). Second, the computation of the inverse covariance $(\mathbf{D} - \mathbf{W}) \otimes \bm{\Sigma}^{-1}$ requires only the inverse of an $I \times I$ matrix, meaning posterior exploration can be done much quicker with MCAR priors than with other competing areal models that require larger matrix inversions. {An MCAR prior also offers a richer correlation structure than is provided by a set of $I$ independent, univariate CAR models, since it captures spatial cross-correlations between pairs of the $I$ measurements through the off-diagonal parameters in $\bm{\Sigma}$. In this sense, MCAR priors offer two advantages. First, as we illustrate for hurricane damage, these cross-correlations have nice interpretations for many natural phenomena, meaning posterior inference on $\boldsymbol{\Sigma}$ provides valuable insight on the spatial tendencies of the phenomenon itself. Second, when such cross-correlations are indeed present in data, accounting for it allows spatial information to be shared between different measurements, which can then provide lower prediction errors compared to independent CAR priors. This is illustrated in our simulations in Section 5.}\\

{However, a problem with \eqref{eq:MCARjoint2} is the singularity of $\mathbf{D} - \mathbf{W}$, which implies that the system is over-parametrized. To maintain identifiability, one solution is to impose zero-sum constraints on the vectors $\boldsymbol{\gamma}_{i,\cdot} = (\gamma_{i,1}, \cdots, \gamma_{i,S})^T, \; i = 1, \cdots, I$ (e.g., Chapter 3 in \citealp{BCG2004}), giving the so-called \textit{intrinsic} MCAR prior. Letting $\lambda_1$ and $\lambda_S$ be the minimum and maximum eigenvalue of $\mathbf{D}^{-1/2}\mathbf{W}\mathbf{D}^{-1/2}$, respectively, another alternative is to add a propriety parameter $\rho \in (1/\lambda_1, 1/\lambda_S)$ to make $\mathbf{D} - \rho \mathbf{W}$ invertible, giving the \textit{proper} MCAR prior. For posterior exploration and prediction, we choose intrinsic priors for two reasons. First, the intrinsic form allows for improved smoothing in locations with few observations \citep{Nea2014}, which is important for our application since several locations indeed have few observed storms. Second, as noted in \cite{Pac2009}, the intrinsic MCAR prior gives an intuitive interpretation of each spatial measurement conditionally centered around its neighbours' averages, whereas the proper prior is both difficult to interpret and yields spatial structures that are difficult to justify as a prior (\citealp{Wal2004}). We therefore only use the proper MCAR prior for generating data in our simulation study, since it requires an invertible covariance matrix.}\\

This framework can be extended to model binary variables, and forms the basis of the storm path model in Section 4. Let $\delta_s$ be an indicator variable for an event at location $s$ (with $\delta_s = 1$ if the event occurs and $0$ otherwise), and let $\boldsymbol{\delta} = (\delta_1, \cdots, \delta_S)^T$. An \textit{autologistic CAR model}, denoted as ${AL(\boldsymbol{\theta},\phi)}$, is given by the conditional specification
\begin{align}
\label{eq:ALfullcond}
\begin{split}
\textrm{logit } \left\{P(\delta_s = 1\; | \; \boldsymbol{\delta}_{(-s)})\right\}  = \bm{x}_s^T \boldsymbol{\theta} + \phi \sum\limits_{r \sim s}\delta_r,
\end{split}
\end{align}
where $\boldsymbol{\delta}_{(-s)}\equiv \{\boldsymbol{\delta}_{r} : r \neq s\}$, and $\bm{x}_s$ is the covariate vector at location $s$ with coefficients $\boldsymbol{\theta}$. In our storm path model, indicators for each location will be used as covariates to model both spatial main effects and ENSO-spatial interactions. The parameter $\phi > 0$ quantifies the clustering tendency of events; a larger $\phi$ implies a higher probability for an event occurring at location $s$ given the same event occurs at its neighbours. Using Brook's Lemma again, the joint distribution of \eqref{eq:ALfullcond} becomes
\begin{equation}
p( \boldsymbol{\delta}) = \frac{\textrm{exp } \left\{\boldsymbol{\theta}^T \left(\sum\limits_{s=1}^S \delta_s \bm{x}_s \right) + \phi \sum\limits_{r \sim s, r<s} \delta_r \delta_s \right\}}{\sum\limits_{\boldsymbol{c} \in \{0,1\}^S} \textrm{exp } \left\{\boldsymbol{\theta}^T \left(\sum\limits_{s = 1}^S c_s \bm{x}_s \right) + \phi \sum\limits_{r \sim s, r < s} c_r c_s \right\}}, \quad \boldsymbol{\delta} \in \{0,1\}^S.
\label{eq:ALjointnormed}
\end{equation}


\section{New model: Regional HTS damage prediction}

The proposed regional prediction model has three parts: storm count, path and damage-per-storm. Only ENSO, seasonal and spatial effects are considered here (since there are limited storm data), but it is straightforward to generalize the framework to include other predictors as well.

\subsection{Model specification}
The first component of the proposed model looks at storm counts. Similar to \cite{LZ2012}, the counts process $\{N(t), t \geq 0\}$ is modelled as a non-homogenous Poisson process (PP) with intensity $\lambda (t)$, where $t$ measures time in years. Let $k(t)$ denote the ENSO phase at time $t \geq 0$, with $k(t) = 1$ indicating El Ni\~no, 2 Neutral, and 3 La Ni\~na. In Section 2, we observed that both ENSO effects and ENSO-seasonal interactions are potentially important, therefore we propose the following model for $\lambda(t)$:
\begin{equation}
{\log \lambda(t) = \left\{\begin{array}{l l}
\beta^{0}_{k(t)} + \sum_{p=1}^P \left( u_{k(t)}^{(p)} \sin\left[\frac{2\pi p}{7/12}\left(t-\frac{4}{12}\right)\right] + v_{k(t)}^{(p)} \cos\left[\frac{2\pi p}{7/12}\left(t-\frac{4}{12}\right)\right] \right), & \textrm{if $\frac{4}{12} < t \leq \frac{11}{12}$},\\
-\infty, & \textrm{otherwise.}
\end{array}\right.}
\label{eq:logint}
\end{equation}
The above specification admits a different regression model for $\log \lambda(t)$ in each ENSO phase, with $\beta^0_k$ as the intercept, and {$u^{(p)}_{k}$ and $v^{(p)}_{k}$ parametrizing the amplitude of sinusoidal waves with frequencies $\frac{p}{7/12}$ in ENSO phase $k$, up to a total of $P$ frequencies.} Similar formulations can be found in \citealp{Sea1999} and \citealp{Pea2005}. This allows us to model not only the ENSO main effects (i.e., varying count frequencies by ENSO) through different choices of $\beta^0_k$, but also ENSO-seasonal interactions (i.e., varying seasonal patterns by ENSO) through different $u^{(p)}_k$ and $v^{(p)}_k$. Adjusting for these different seasonal patterns is important because it allows the proposed model to account for the lower concentration of storms (and hence storm damage) in ASO for Neutral years (see Figure \ref{fig:histpath}). {Since all storms occurred between the months of May to November inclusive, $\lambda(t)$ is restricted to be positive only within that time period, and zero otherwise.} A similar model was proposed in \cite{Cea2011}, however, a key difference is that we model annual seasonality within each ENSO phase.\\

The second component of our model considers storm paths. Define $G = (V,E)$ as the (undirected) graph with $V$ as the set of locations and $E$ as the set of edges under LOS connectivity. Let $\boldsymbol{\delta} = (\delta_1, \delta_2, \cdots, \delta_{S})^T$ be the indicator vector for a given storm (with $\delta_s = 1$ if the storm hits location $s$, and 0 otherwise), and let $\gamma_{k,s}$ be the spatial effect parameter for location $s$ in ENSO phase $k$. Given a storm count from $\{N(t), t \geq 0\}$ in ENSO phase $k$, its path $\boldsymbol{\delta}$ is modelled as
\begin{align}
\label{eq:AppALcond}
\begin{split}
\textrm{logit } \left\{P(\delta_s = 1 \; | \; \boldsymbol{\delta}_{(-s)})\; \right\} = \gamma^0_k + \gamma_{k,s} + \phi \sum\limits_{r \sim s}\delta_r, \quad \forall s = 1, \cdots, S,
\end{split}
\end{align} 
where $\sim$ is the adjacency operator on $G$. Each storm path is also assumed to be independent of each other and conditionally independent of $\{N(t), t \geq 0\}$. In light of \eqref{eq:ALfullcond}, the specification in \eqref{eq:AppALcond} allows for different autologistic models in each ENSO phase $k$, with $\gamma_k^0$ as the intercept and $\boldsymbol{\gamma}_{k,\cdot} = (\gamma_{k,1}, \gamma_{k,2}, \cdots, \gamma_{k,S})^T$ as the vector of spatial effects. This framework again allows us to model the ENSO main effects (i.e., varying number of locations hit per storm by ENSO) through $\gamma^0_k$, as well as ENSO-spatial interactions (i.e., varying spatial patterns of paths by ENSO) through $\boldsymbol{\gamma}_{k,\cdot}$. To maintain parameter identifiability, we enforce the zero-sum constraints $\sum_{s=1}^S {\gamma}_{k,s} = 0, \; k = 1,2,3$ \citep{BCG2004}. The parameter $\phi > 0$ in \eqref{eq:AppALcond} again quantifies storm clustering tendencies; a larger $\phi$ increases the chance of a location being hit given its neighbours are also hit. \\

Although Brook's Lemma can be applied to \eqref{eq:AppALcond} to get the likelihood \eqref{eq:ALjointnormed}, we do not recommend doing so, since it assigns positive probability to storm paths that cannot occur (i.e., LOS disconnected paths). Put another way, since all historical storms are LOS connected, allowing for disconnected path predictions results in geographically impossible forecasts. To rectify this, we sum over only LOS connected paths in the denominator of \eqref{eq:ALjointnormed}, and assign zero probability to LOS disconnected paths (including the null path $\bm{\delta} = (0, \cdots, 0)^T$).\\

The final component of our model considers regional damage in each location from a given storm. Suppose the $i$-th storm occurs in ENSO phase $k$ with path $\boldsymbol{\delta}^{(i)} = ({\delta}_1^{(i)}, \cdots, {\delta}_S^{(i)})^T$, and let $Y^{(i)}_{s}$ be the damage incurred at location $s$. If the storm does not hit location $s$ (i.e., $\delta_s^{(i)} = 0$), then clearly $Y^{(i)}_{s}=0$. However, if $s$ is indeed hit (i.e., $\delta_s^{(i)} = 1$), then, following \cite{Kat2002}, we model $Y_s^{(i)}$ by the lognormal model
\begin{align}
\label{eq:myLNmodel}
Y^{(i)}_{s} | \boldsymbol{\delta}^{(i)} \distas{indep.}\; \textrm{Lognormal}\left({\xi^0_k} + {\xi}_{k,s} + \zeta^{(i)}, \sigma^2 \right), \quad \forall s: \delta_s^{(i)} = 1,
\end{align}
with \eqref{eq:myLNmodel} conditionally independent of both $\{N(t),t \geq 0\}$ and $\boldsymbol{\delta}_{-s}^{(i)}$. The specification in \eqref{eq:myLNmodel} allows for different lognormal models in each ENSO phase, and, as before, ENSO main effects and ENSO-spatial interactions can be modelled by varying $\xi^0_{k}$ and $\boldsymbol{\xi}_{k,\cdot} = (\xi_{k,1}, \xi_{k,2}, \cdots, \xi_{k,S})^T$. Zero-sum constraints $\sum_{s=1}^S {\xi}_{k,s} = 0, \; k = 1, 2, 3$ are again enforced for identifiability. The random effect $\zeta^{(i)}$ models storm-to-storm variability for damage (call these \textit{storm severity} effects); larger values of $\zeta^{(i)}$ indicate above-average damage in all affected locations along the path for storm $i$, and vice versa. We assume that these random effects are i.i.d. $N(0,\sigma^2_\zeta)$.\\

Lastly, the proposed process for $X_s(t)$, the HTS damage incurred in the time interval $[0,t]$ at location $s$, is obtained by combining the three submodels:
\begin{equation}
X_s(t) = \sum_{i = 1}^{N(t)} \left( \mathbf{1}_{\{\delta^{(i)}_s = 1\}} \cdot Y^{(i)}_s \right), \quad s = 1, \cdots, S, 
\label{eq:Xtagg}
\end{equation}
where $\{N(t), t \geq 0\}$, $\boldsymbol{\delta}^{(i)}|\{N(t)\}$ and $Y_s^{(i)}|\boldsymbol{\delta}^{(i)}, \{N(t)\}$ are mutually independent. This framework is similar in structure to the compound Poisson-lognormal model \eqref{eq:CPL}, with the important distinction that \eqref{eq:Xtagg} allows for regional analysis and damage prediction. Note that, for LOS connected locations $r$ and $s$, $X_r(t)$ and $X_s(t)$ are correlated through both the storm path and damage submodels. This correlation structure allows for forecasts to be made in locations with little to no observations by borrowing information from LOS neighbours.

\begin{table}
\ttabbox{
\centering
\begin{tabular}{ l |  l  l }
	\toprule
	\multicolumn{1}{c}{\textit{Model}} & \multicolumn{2}{c}{\textit{Distribution}} \\
	\toprule
	\textbf{Count} &  $[N(t) | \lambda(t)]$ & $N(t) \sim $ Pois$(\lambda(t))$\\
	Intensity & $[\lambda(t)|\boldsymbol{\beta}^0, \boldsymbol{u}, \boldsymbol{v}]$ & See \eqref{eq:logint}\\
	Priors: & $[\boldsymbol{\beta}^0, \boldsymbol{u}, \boldsymbol{v}] = [\boldsymbol{\beta}^0][\boldsymbol{u}][\boldsymbol{v}]$ &\\
	\qquad {\textit{ENSO}} & & $[\boldsymbol{\beta}^0] \propto 1$\\
	\qquad {\textit{Seasonality}} & & $[\boldsymbol{u}] \propto 1$, $[\boldsymbol{v}] \propto 1$\\
	\toprule
	\textbf{Path} & $[\boldsymbol{\delta}^{(i)}| \boldsymbol{\gamma}^0, \boldsymbol{\gamma}, \phi, \boldsymbol{\Sigma}_{\gamma}]$ & See \eqref{eq:AppALcond}\\
	Priors: & $[\boldsymbol{\gamma}^0, \boldsymbol{\gamma}, \phi | \bm{\Sigma}_{\gamma}] =  [\boldsymbol{\gamma}^0] [\boldsymbol{\gamma}| {\Sigma}_{\gamma}] [\phi]$ & \\
	\qquad {\textit{ENSO}} & & $[\boldsymbol{\gamma}^0] \propto 1$\\
	\qquad {\textit{Spatial}} & & $[\boldsymbol{\gamma} | \bm{\Sigma}_{\gamma}] \sim MCAR(\bm{\Sigma}_{\gamma})$\\
	\qquad {\textit{Clustering}} & & $[\phi] \sim IG(0.01,0.01)$\\
	Hyperpriors: & $[\bm{\Sigma}_{\gamma}]$ & $[\bm{\Sigma}_{\gamma}] \sim IW(4,\mathbf{I})$\\
	\toprule
	\textbf{Damage} & $[Y_s^{(i)}|\boldsymbol{\xi}^0, \boldsymbol{\xi}, \zeta, \sigma^2, \sigma^2_\zeta, \bm{\Sigma}_\xi]$ & See \eqref{eq:myLNmodel}\\
	Priors: & $[\boldsymbol{\xi}^0, \boldsymbol{\xi},\zeta, \sigma^2|\sigma^2_{\zeta},\bm{\Sigma}_{\xi}] = [\boldsymbol{\xi}^0] [\boldsymbol{\xi}|\bm{\Sigma}_{\xi}][\zeta|\sigma^2_{\zeta}] [\sigma^2]$ & \\
	\qquad \textit{ENSO} & & $[\boldsymbol{\xi}^0] \propto 1$\\
	\qquad {\textit{Spatial}} & & $[\boldsymbol{\xi}|\bm{\Sigma}_{\xi}] \sim MCAR(\bm{\Sigma}_{\xi})$\\
	\qquad {\textit{Severity}} & &$[\zeta|\sigma^2_{\zeta}] \sim N(0, \sigma^2_\zeta)$\\
	\qquad {\textit{Log-damage variance}} & &$[\sigma^2] \sim IG(0.01,0.01)$\\
	Hyperpriors: & $[\bm{\Sigma}_{\xi}]$ & $[\bm{\Sigma}_{\xi}] \sim IW(4, \mathbf{I})$ \\
	& $[\sigma^2_{\zeta}]$ & $[\sigma^2_{\zeta}] \sim IG(0.01,0.01)$ \\
	\toprule
\end{tabular}
}
{\caption{Hierarchical summary for storm count, path and damage-per-storm submodels.}
\label{table:hiermodel}
}
\end{table}

\subsection{Prior distributions}
For statistical inference, we take a Bayesian approach. In some applications, information may exist to guide the selection of prior distributions. Lacking such information, weakly informative or non-informative priors are used instead for parameters in all three submodels. For the spatial effect vectors $\boldsymbol{\gamma}= (\boldsymbol{\gamma}_{\cdot,1}^T, \cdots, \boldsymbol{\gamma}_{\cdot,S}^T)^T$ and $\boldsymbol{\xi} = (\boldsymbol{\xi}_{\cdot,1}^T, \cdots, \boldsymbol{\xi}_{\cdot,S}^T)^T$, we assign the Gaussian CAR priors $MCAR(\bm{\Sigma}_\gamma)$ and $MCAR(\bm{\Sigma}_\xi)$, respectively. These priors provide a flexible framework for modeling a wide range of spatial correlation structures. In particular, the diagonal entries in $\bm{\Sigma}_{\gamma}$ and $\bm{\Sigma}_{\xi}$ account for spatial variation within each ENSO phase, whereas off-diagonal entries control spatial correlation between different pairs of phases. {For example, $\bm{\Sigma}_{\gamma,1,2} > 0$ and $\bm{\Sigma}_{\xi,1,2} > 0$ indicate that, after accounting for ENSO main effects, the spatial tendencies of storm paths and damage are positively correlated in El Ni\~no and Neutral years.} Posterior exploration of $\bm{\Sigma}_\gamma$ and $\bm{\Sigma}_\xi$ are therefore not only useful for prediction, but also for gaining a deeper insight on how ENSO influences spatial patterns for both storm path and damage.\\

As for the remaining model parameters, we assign to each intercept term ($\boldsymbol{\beta}^0 = (\beta_1^0, \beta_2^0, \beta_3^0)^T$ for counts, $\boldsymbol{\gamma}^0 = (\gamma_1^0, \gamma_2^0, \gamma_3^0)^T$ for path, and $\boldsymbol{\xi}^0 =  (\xi_1^0, \xi_2^0, \xi_3^0)^T$ for damage) and seasonality parameter ($\boldsymbol{u} = (\boldsymbol{u}_{\cdot,1}^T, \cdots, \boldsymbol{u}_{\cdot,P}^T)^T$ and $\boldsymbol{v} =  (\boldsymbol{v}_{\cdot,1}^T, \cdots, \boldsymbol{v}_{\cdot,P}^T)^T$, where $\boldsymbol{u}_{\cdot,p} = (u_1^{(p)},u_2^{(p)},u_3^{(p)})$ and $\boldsymbol{v}_{\cdot,p} = (v_1^{(p)},v_2^{(p)},v_3^{(p)})$) improper priors on $\mathbb{R}$. To maintain parameter identifiability, $\sigma^2$, $\sigma^2_{\zeta}$ and $\phi$ are assigned weak inverse-gamma priors $IG(0.01,0.01)$. The covariance matrices $\bm{\Sigma}_{\gamma}$ and $\bm{\Sigma}_{\xi}$ are assigned {conjugate} inverse-Wishart hyperpriors $IW(4, \mathbf{I})$, following Section 3.6 in \cite{Gea2014}. Table \ref{table:hiermodel} provides a concise summary of this prior specification.

\subsection{Posterior exploration}

For posterior exploration, we use an MCMC hybrid sampler proposed by \cite{Tie1994}, which uses full conditional updates whenever possible (for efficient MCMC mixing) and Metropolis updates \citep{Mea1953} otherwise. Since the three submodels are conditionally independent, separate samplers can be used for each submodel. {For the counts submodel, we use the exact likelihood expression for the non-homogenous PP provided in \cite{DV2003}, and approximate its integrals by numerical integration. However, under this likelihood, all parameters have non-standard conditional distributions, so the full Metropolis-within-Gibbs sampler \citep{Mea1953} is employed.} We briefly describe the samplers for path and damage below, with derivations found in Appendix B.\\

Let $\Theta^P = \{\boldsymbol{\gamma}^0, \boldsymbol{\gamma}, \phi, \bm{\Sigma}_\gamma\}$ be the parameter set for the path submodel. After tuning for optimal step-sizes (described later in section) and setting initial parameter values, the path sampler iterates through the following steps:	
\bi
\item Use Metropolis updates for $\phi$ and each parameter in $\boldsymbol{\gamma}^0$ and $\boldsymbol{\gamma}$,
\item For $k = 1, 2, 3$, apply zero-sum constraints by centering each $\boldsymbol{\gamma}_{k,\cdot}$ by its mean, as suggested in \cite{BCG2004}, and
\item Update $\bm{\Sigma}_\gamma$ by its full conditional \eqref{eq:cond1}.
\ei
{Although sampling $\boldsymbol{\gamma}_{k,\cdot}$ under this ``on-the-fly'' correction of centering by its mean is not equivalent to sampling directly under the zero-sum constraint \citep{RH2005}, the difference is minimal in practice \citep{Pac2009}.}\\

{Likewise, let $\Theta^D = \{\boldsymbol{\xi}^0, \boldsymbol{\xi}, \sigma^2, \sigma^2_\zeta, \bm{\Sigma}_\xi\}$ be the parameter set for the damage submodel. Since $\zeta^{(i)}$ (storm severity effect) and $Y_s^{(i)}$ (regional damage of storm $i$ at location $s$) are unobserved, both need to be sampled along with the parameters in $\Theta^D$. There are two difficulties in sampling the latter. First, the distribution of independent lognormal random variables conditional on its sum is non-standard (unlike for the gamma distribution, which is proportional to a Dirichlet distribution). Second, the distribution of the sum of lognormals is non-standard as well. To this end, we use an approximation scheme proposed in \cite{Mea2007} that approximates the sum of lognormals as a new lognormal random variable by matching their respective moment-generating function (mgf) at specific points. Using this approximation, and letting $Y^{(i)}$ be the \textit{observed} total damage from storm $i$, the conditional regional damage $\{ Y_s^{(i)} : \delta^{(i)}_s=1\} | Y^{(i)}$ can then be sampled with Metropolis updates. Details for this sampling scheme are provided in Appendix C.}\\

After tuning step-sizes and initializing parameter values, the damage sampler cycles through the following steps:
\bi
\item Use Metropolis steps for each parameter in $\boldsymbol{\xi}^0$ and $\boldsymbol{\xi}$,
\item For $k = 1, 2, 3$, apply zero-sum constraints by centering each $\boldsymbol{\xi}_{k,\cdot}$ by its mean,
\item Update $\bm{\Sigma}_\xi$ and $\sigma^2_\zeta$ by its full conditionals \eqref{eq:cond2} and \eqref{eq:cond3},
\item Sample $\zeta^{(i)}$ from \eqref{eq:cond5} for all storms $i = 1, \cdots, n$ in data, and
\item {For each multiple-location storm, sample $\{ Y_s^{(i)} : \delta^{(i)}_s=1\} | Y^{(i)}$ using the sampling scheme in Appendix C.}
\ei

{For Metropolis steps, a Gaussian random walk proposal distribution is used with step-size equal to its standard deviation.} The choice of this step-size is crucial for efficient MCMC mixing, since it allows the sampler to sufficiently explore the parameter space in a reasonable amount of time. An acceptance rate of around 44\% is recommended for optimal mixing of univariate MCMC \citep{Rea1997}, so, to achieve roughly this benchmark for each update in the sampler, we tune step-sizes using the algorithm in \cite{Gra2011}. {To foreshadow, the acceptance rates in both our simulation and actual analysis ranged from 17\% to 49\%, and, using the convergence diagnostics described below, the sampled chains show adequate mixing as well.}\\

To assess MCMC convergence, we use trace plot inspection and the Brooks-Gelman-Rubin (BGR) statistic \citep{Gea2014}, which reports a ratio of within- and between-chain variances (with values much larger than 1 suggesting poor mixing). Close attention is paid to these, since they not only help indicate MCMC convergence, but also help identify parameter identifiability issues \citep{Nea2014}.


\section{Simulation study}

We have three goals for our simulation study: to explore MCMC {convergence}, to ensure good parameter inference, and to assess how well our model provides regional inference and predictions. The last goal is particularly important, since it shows the proposed model can indeed give good regional forecasts with only aggregated damage data. To investigate these goals, several storm datasets were simulated, each incorporating different combinations of ENSO, spatial and seasonal effects for counts, path and damage. Conclusions are consistent for all simulation studies, so, for brevity, we picked only one study to discuss. \\

\begin{figure}
\begin{floatrow}
\capbtabbox{
\small
\begin{tabular}{c | c }
\toprule
{\textit{Parameters}} & {\textit{Avg. posterior means}}\\
\toprule
$\boldsymbol{\beta}^0 = (1.75,2,2.25)$ & $(1.75, 1.97, 2.22)$ \\
$\boldsymbol{u} = (0,0.5,-0.5) $ & $(-0.01, 0.51, -0.51)$\\
$\boldsymbol{v} = (0,0.5,-0.5)$ & $(-0.01, 0.52, -0.50)$\\
\hline
$\boldsymbol{\gamma}^0 = (-4.25,-4,-3.75) $ & $(-4.22, -3.90, -3.61)$\\
$\phi = 1$ & 0.93\\
$\bm{\Sigma}_\gamma = 0.2\cdot\bm{I} + 0.8 \cdot \bm{J}$ & $\begin{pmatrix} 1.12 & 0.84 & 0.83 \\ 0.84 & 1.15 & 0.85 \\ 0.83 & 0.85 & 1.14 \end{pmatrix}$\\
\hline
$\boldsymbol{\xi}^0 = (18,20,22)$ & $(17.8, 20.1, 22.0)$\\
$\sigma^2 = 5$ & 5.24 \\
$\sigma^2_\zeta = 1$ & 1.07 \\
$\bm{\Sigma}_\xi = 0.8\cdot \bm{I} + 0.2 \cdot \bm{J}$ & $\begin{pmatrix} 1.32 & 0.08 & 0.10 \\ 0.08 & 1.14 & 0.14 \\ 0.10 & 0.14 & 1.06 \end{pmatrix}$ \\
\toprule
\end{tabular}
}
{\caption{Parameter choices and {average posterior means} for our simulation study. $\mathbf{J}$ is a matrix of ones.}
\label{tbl:sim}
}
\hspace{0.1cm}
\ffigbox{\includegraphics[scale=0.55]{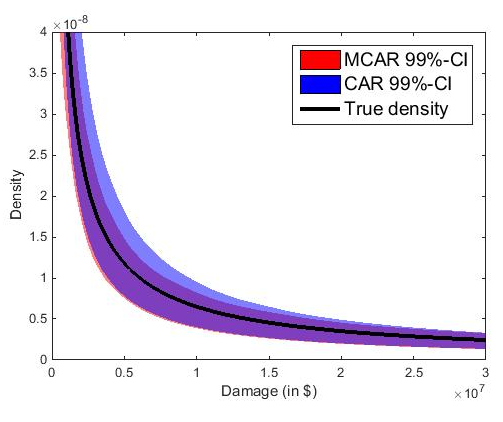}}
{\caption{True and pointwise 99\% HPD bands (red for MCAR prior, blue for CAR priors) of the lognormal damage density for AL in a Neutral year. }
\label{fig:ci}
}

\end{floatrow}
\end{figure}

The chosen study simulates storm data exhibiting ENSO-seasonal interactions for counts (with $P=1$), as well as ENSO-spatial interactions for both storm path and damage. The left column in Table \ref{tbl:sim} summarizes the choice of parameters used for simulation. Each dataset is obtained by first simulating storm counts from \eqref{eq:logint}, then simulating its corresponding path and damage from \eqref{eq:AppALcond} and \eqref{eq:myLNmodel}, respectively. As mentioned in Section 3, to simulate the spatial effects $\boldsymbol{\gamma}$ and $\boldsymbol{\xi}$ from \eqref{eq:MCARjoint2}, a parameter $\rho = 0.99$ is added to make $\mathbf{D} - \rho\mathbf{W}$ invertible (this is in contrast to posterior exploration, where the intrinsic form $\mathbf{D} - \mathbf{W}$ is used). To replicate conditions in the original data, the same set of locations and El Ni\~no, Neutral and La Ni\~na years are used. 20,000 MCMC iterations are run for each simulation, with the first 5,000 discarded as burn-in and the remaining samples thinned by keeping every 10-th sample. This procedure is then replicated 100 times to compute average posterior means and coverage rates for highest posterior density (HPD) intervals. \\

{For the first goal, trace plot inspection and BGR statistics show adequate MCMC convergence for a variety of effect combinations}. {For the second goal, the right part of Table \ref{tbl:sim} summarizes the average posterior means over all simulations. These posterior means can be seen to be quite close to true values, and coverage rates for 95\% HPD intervals (not shown here) are above 90\% for all parameters, indicating that inference for the proposed model (when equipped with weakly-informative priors) performs quite well. This gives confidence that our method can indeed detect effects that are present in the data generating mechanism.} Lastly, to assess regional performance, {Figure \ref{fig:ci} compares the true damage density from \eqref{eq:myLNmodel} and its 99\% pointwise HPD bands for a select location (AL) in a randomly chosen simulation, ignoring storm severity. The red band corresponds to the proposed model, which uses an MCAR prior for $\boldsymbol{\Sigma}_\xi$, and the blue band corresponds to the model fit using three independent CAR priors. Two interesting observations can be seen. First, the true density is fully contained in the red (MCAR) confidence band, which suggests the proposed model gives good regional inference for damage despite fitting with aggregate data. Second, although both the red and blue bands cover the true density, the latter is noticeably wider than the former. This illustrates an advantage of using MCAR priors: By allowing spatial information to be shared between ENSO phases, the MCAR prior {can} provide higher precision of estimates compared to independent CAR priors {when non-diagonal  $\Sigma_{\gamma}$ and $\Sigma_{\xi}$ are used to generate the data. This non-diagonality assumption is not unreasonable to make for our application, since, intuitively, one expects locations which are hit often or incur higher damage in one ENSO phase to also be hit often or incur higher damage in other phases as well. For these reasons, we explore only the MCAR model in the following analysis of the HTS data.} }

\floatsetup[figure]{style=Boxed,frameset={\fboxrule1pt}}
\floatsetup[figure]{style=plain}

\section{Analysis of HTS data}

\subsection{MCMC results and predictions}

\floatsetup[figure]{style=Boxed,frameset={\fboxrule1pt}}
\begin{figure}[t]
\begin{floatrow}
\centering
\ffigbox{\includegraphics[scale=0.37]{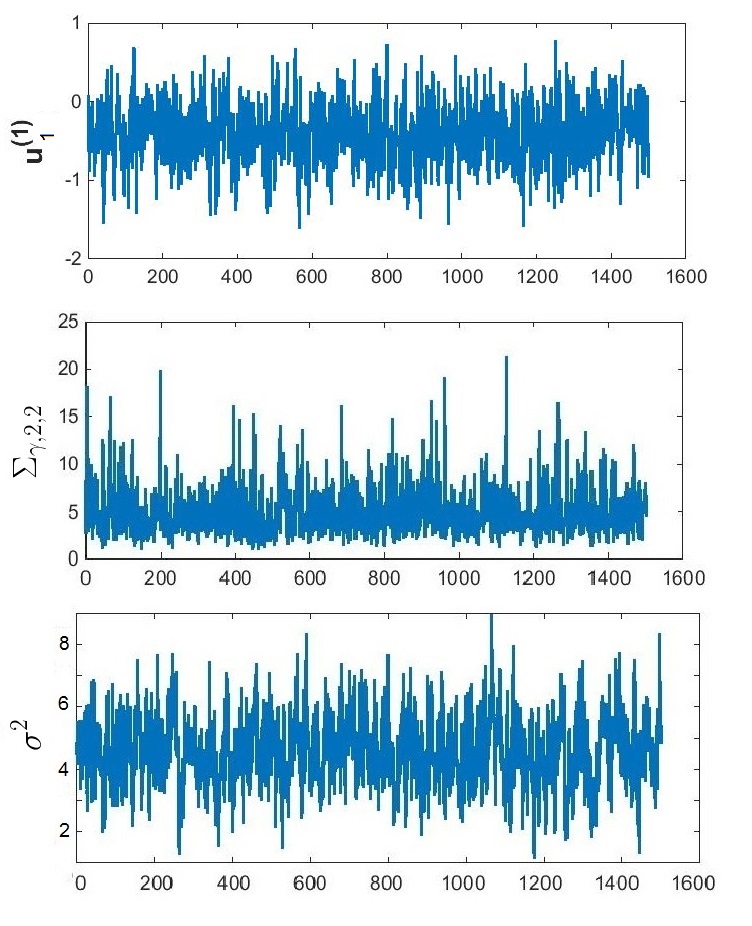} \vspace{0.5cm}}
{\caption{Posterior trace plots of $u^{(1)}_1$ (BGR statistic: 1.0074) from counts model, $\Sigma_{\gamma,2,2}$ (1.0122) from paths model, and $\sigma^2$ (0.9982) from damage model.}
\label{fig:trace}
}
\hspace{0.1cm}
\ffigbox{\includegraphics[scale=0.55]{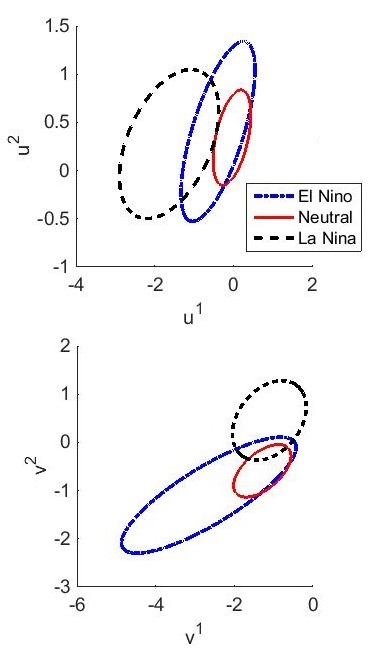}}
{\caption{95\% HPD regions for the first two frequency coefficients.}
\label{fig:hpdregs}
}
\end{floatrow}
\end{figure} 
\floatsetup[figure]{style=plain}

We now fit the proposed model to the historical HTS data in Section 2. For each submodel, 20,000 iterations of the MCMC sampler are run, with the first 5,000 removed as burn-in. To reduce autocorrelation, the remaining MCMC samples are thinned by keeping every 10-th sample. Figure \ref{fig:trace} provides the trace plots (after burn-in and thinning) for a select parameter from each submodel. These plots provide no visual indication of nonstationarity and show little autocorrelation as well, which is as desired. This convergence is also confirmed by the BGR statistics, which are all sufficiently close to 1. Similar conclusions hold for remaining parameters as well. Table \ref{tbl:post} summarizes the posterior mean and 95\% HPD interval for each parameter.\\

\begin{figure}[t]
\centering
\centering
\includegraphics[scale=0.5]{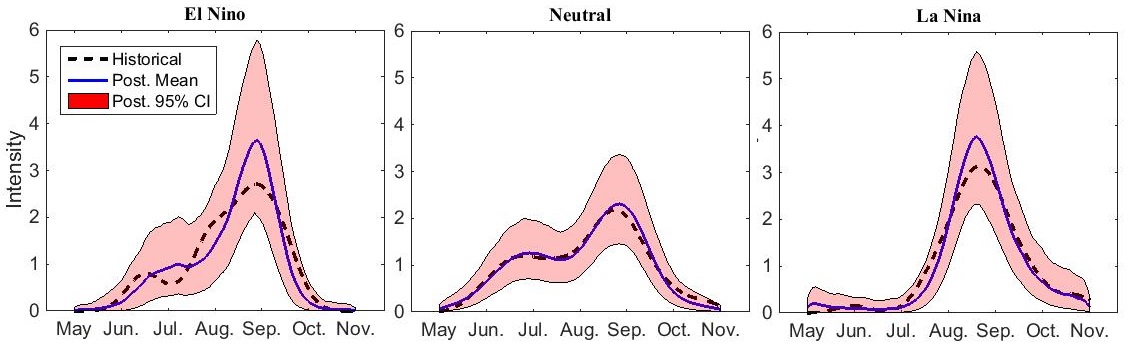}
\caption{Estimated historical and posterior mean count intensities for each ENSO phase. The red region marks the 95\% pointwise HPD region.}
\label{fig:simCounts}
\end{figure}

{For storm counts, the posterior means and 95\% HPD intervals of expected storms in an El Ni\~no, Neutral and La Ni\~na year are 2.06 [1.44, 2.82], 2.44 [1.92, 3.08] and 3.31 [2.37, 4.30], respectively, which confirms previous findings that cooler anomalies are associated with higher storm counts. As for total frequencies $P$ to include, we fit the counts submodel with $P = 1, \cdots, 5$ and, using the Deviance Information Criterion (DIC, \citealp{Sea2002}), found that $P=2$ frequencies provide the best fit for the data. To investigate whether there is evidence for ENSO-seasonal interactions, Figure \ref{fig:hpdregs} plots the 95\% HPD regions of $(u^{(1)}, u^{(2)})$ and $(v^{(1)}, v^{(2)})$ for each ENSO phase. The regions corresponding to Neutral and La Ni\~na years experience little-to-no overlaps, which suggests that seasonal patterns for these two phases are significantly different, and provides evidence for ENSO-seasonal interactions. On the other hand, the HPD region for Neutral years is nearly contained within that for El Ni\~no, indicating that seasonality for these two phases are quite similar. These results can be seen more clearly in Figure \ref{fig:simCounts}, which plots the kernel approximation of historical count intensities with the posterior mean intensities and its 95\% pointwise HPD intervals. For El Ni\~no and Neutral years, two intensity peaks can be observed: a large peak at the end of August, and a smaller one at the end of June, which is in line with the two-frequency model selected by DIC. The seasonal pattern in La Ni\~na, however, is visually quite different from the other two ENSO phases, in that it exhibits only the larger peak in August.} \\

{As for storm paths, $\hat{\gamma}^0_1 = -4.83$ is larger than $\hat{\gamma}^0_2 = -5.68$ and $\hat{\gamma}^0_3 = -5.91$, where, for the following discussion, $\hat{\gamma}$ denotes the estimated posterior mean of parameter $\gamma$.} This suggests that more locations tend to be hit in an El Ni\~no storm than in a Neutral or La Ni\~na storm. A strong storm clustering tendency is also implied by the clustering parameter estimate $\hat{\phi} = 1.49$. For spatial variability of storm paths, $\hat{\bm{\Sigma}}_{\gamma,1,1} = 2.85$, $\hat{\bm{\Sigma}}_{\gamma,2,2} = 5.97$ and $\hat{\bm{\Sigma}}_{\gamma,3,3} = 7.30$, which indicates that, although there are noticeable spatial patterns for each ENSO phase, these effects are more pronounced in Neutral and La Ni\~na years than in El Ni\~no. {The posterior means for spatial correlations are also quite large, with $\hat{\rho}_{\gamma,1,2} = 0.89 \; [0.71, 0.98]$, $\hat{\rho}_{\gamma,1,3} = 0.89\; [0.70,0.98]$ and $\hat{\rho}_{\gamma,2,3}= 0.95 \;[0.87, 0.99]$.} This indicates that locations frequently hit in one ENSO phase tend to be frequently hit in other phases as well. \\

For storm damage, storms in El Ni\~no years incur notably less damage ($\hat{\xi}^0_1 = 18.60$) than Neutral or La Ni\~na storms ($\hat{\xi}^0_{2} = 19.86$, $\hat{\xi}^0_{3} = 19.56$), which is consistent with our analysis in Section 2 and findings in \cite{PL1998}. There is also a noticeable storm severity effect, $\hat{\sigma}^2_\zeta = 1.66$, although this effect is small compared to the log-damage variance $\hat{\sigma}^2 = 4.75$. For spatial variability of damage, El Ni\~no storms experience the largest variation ($\hat{\bm{\Sigma}}_{\xi,1,1} = 1.78$), followed by La Ni\~na ($\hat{\bm{\Sigma}}_{\xi,3,3} = 1.70$) and Neutral storms ($\hat{\bm{\Sigma}}_{\xi,2,2} = 0.75$). {The posterior means and 95\% HPD intervals for spatial correlations are $\hat{\rho}_{\xi,1,2} = -0.07\; [-0.36, 0.20]$, $\hat{\rho}_{\xi,1,3} = -0.17\; [-0.44,0.07]$ and $\hat{\rho}_{\xi,2,3} = 0.09\; [-0.21, 0.32]$, which indicate that the spatial pattern for damage in El Ni\~no storms have a slight negative correlation with that for Neutral and La Ni\~na storms}. This is counter-intuitive, since one expects locations that incur higher damage in, say, an El Ni\~no storm, to incur higher damage in, say, a La Ni\~na storm as well. One plausible explanation is that more densely-populated locations are able to better withstand weaker storms, but incur higher damage for more catastrophic storms, which then creates changing spatial patterns for the higher-damaging storms in La Ni\~na and the lower-damaging ones in El Ni\~no.\\

\begin{table}[t]
\ttabbox{
\centering
\begin{small}
\begin{tabular}{c c c c c}
\toprule
{\textit{Model}} & {\textit{Parameter}} & {\textit{Description}} & {\textit{Posterior mean}} & {\textit{95\% HPD}}\\
\toprule
Count & $\beta^0_{1}$ & $EN$ main effect & -0.57 & [-1.75, 0.27]\\
& $\beta^0_{2}$ & $NE$ main effect & 0.46 & [0.07, 0.81]\\
& $\beta^0_{3}$ & $LN$ main effect & 0.23 & [-0.52, 0.83]\\
& $u_{1}^{(1)}, v_{1}^{(1)}$ &  $EN$ seasonality (1$^{st}$ freq.) & -0.40, -2.60 & [-1.19, 0.36], [-4.5, -1.24]\\
& $u_{2}^{(1)}, v_{2}^{(1)}$ & $NE$ seasonality (1$^{st}$ freq.) & -0.03, -1.29 & [-0.40, 0.36], [-1.92, -0.75]\\
& $u_{3}^{(1)}, v_{3}^{(1)}$ & $LN$ seasonality (1$^{st}$ freq.) & -1.65, -1.12 & [-2.78, -0.72], [-1.97, -0.42]\\
& $u_{1}^{(2)}, v_{1}^{(2)}$ &  $EN$ seasonality (2$^{nd}$ freq.) & 0.40, -1.07 & [-0.36, 1.18], [-2.09, -0.25]\\
& $u_{2}^{(2)}, v_{2}^{(2)}$ & $NE$ seasonality (2$^{nd}$ freq.) & 0.35, -0.59 & [-0.04, 0.75], [-1.05, -0.16]\\
& $u_{3}^{(2)}, v_{3}^{(2)}$ & $LN$ seasonality (2$^{nd}$ freq.) & 0.27, 0.45 & [-0.36, 0.88], [-0.19, 1.14]\\
\hline
Path & $\gamma^0_{1}$ & $EN$ main effect & -4.83 & [-6.28, -3.34]\\
& $\gamma^0_{2}$ & $NE$ main effect & -5.68 & [-6.99, -4.26]\\
& $\gamma^0_{3}$ & $LN$ main effect & -5.91 & [-7.33, -4.39]\\
& $\phi$ & Clustering & 1.49 & [0.48, 2.30]\\
& $\bm{\Sigma}_{\gamma,1,1}$ & Spatial var. - $EN$ & 2.85 & [0.66, 7.57]\\
& $\bm{\Sigma}_{\gamma,1,2}$ & Spatial cov. - $EN$/$NE$ & 3.63 & [1.05, 8.93]\\
& $\bm{\Sigma}_{\gamma,1,3}$ & Spatial cov. - $EN$/$LN$ & 4.00 & [1.11, 9.91]\\
& $\bm{\Sigma}_{\gamma,2,2}$ & Spatial var. - $NE$ & 5.97 & [1.12, 14.13]\\
& $\bm{\Sigma}_{\gamma,2,3}$ & Spatial cov. - $NE$/$LN$ & 6.13 & [2.30, 14.23]\\
& $\bm{\Sigma}_{\gamma,3,3}$ & Spatial var. - $LN$ & 7.30 & [2.36, 18.25]\\
\hline
Damage & $\xi^0_1$& $EN$ main effect & 18.60 & [17.67, 19.50]\\
& $\xi^0_{2}$ & $NE$ main effect& 19.86 & [19.16, 20.53]\\
& $\xi^0_{3}$ & $LN$ main effect& 19.56 & [18.72, 20.44]\\
& $\sigma^2$ & Log-damage var. & 4.75 & [2.22, 7.42]\\
& $\sigma^2_\zeta$ & Storm severity & 1.66 & [0.02, 4.33]\\
& $\bm{\Sigma}_{\xi,1,1}$ & Spatial var. - $EN$& 1.78 & [0.15, 8.68]\\
& $\bm{\Sigma}_{\xi,1,2}$ & Spatial cov. - $EN$/$NE$& -0.14 & [-2.15, 1.46]\\
& $\bm{\Sigma}_{\xi,1,3}$ & Spatial cov. - $EN$/$LN$& -0.47 & [-4.28, 1.70]\\
& $\bm{\Sigma}_{\xi,2,2}$ & Spatial var. - $NE$ & 0.75 & [0.13, 2.78]\\
& $\bm{\Sigma}_{\xi,2,3}$ & Spatial cov. - $NE$/$LN$ & 0.13 & [-1.54, 2.22]\\
& $\bm{\Sigma}_{\xi,3,3}$ & Spatial var. - $LN$ & 1.70 & [0.16, 7.85]\\
\toprule
\end{tabular}
\end{small}
}
{\caption{Posterior means and 95\% HPD intervals for count, path and damage model parameters.}
\label{tbl:post}
}
\end{table}

\begin{figure}[p]
\ffigbox
{\includegraphics[scale = 0.63]{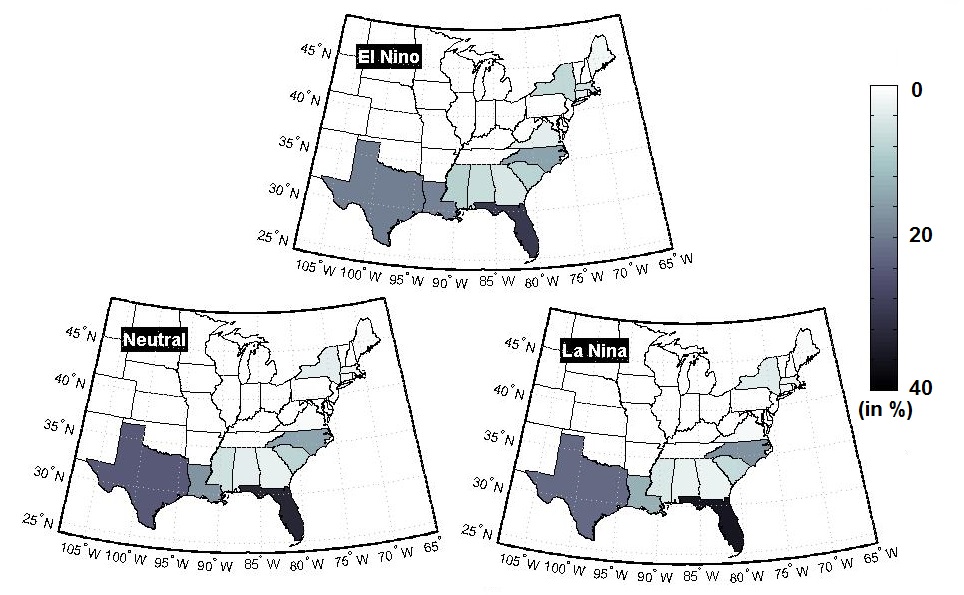}}
{\caption{Simulated hit-rates of storms in each ENSO phase.}
\label{fig:simPath}}
\ffigbox{\includegraphics[scale = 0.63]{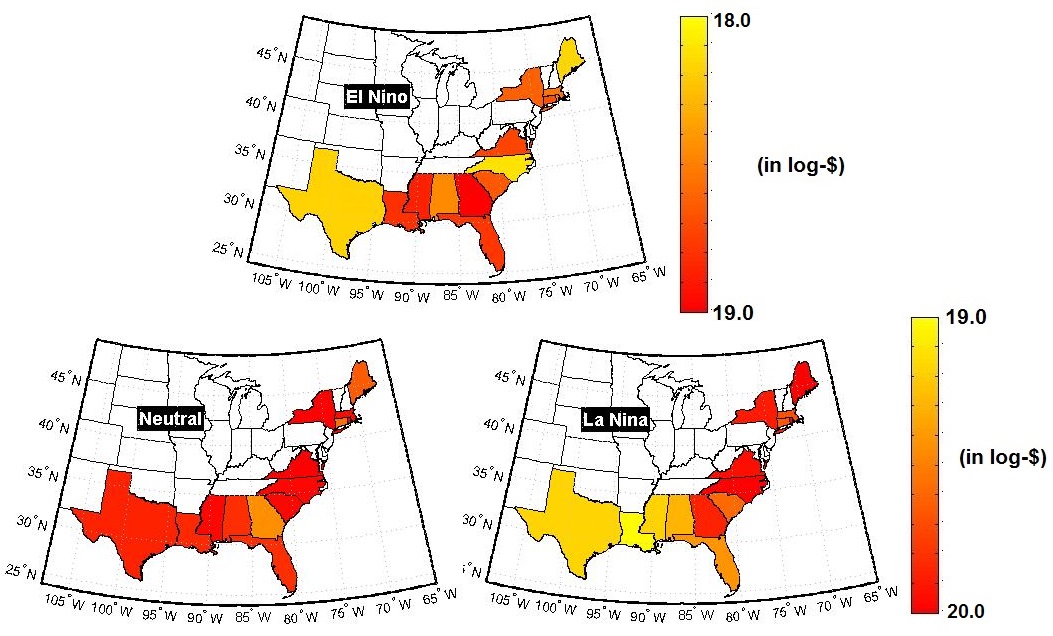}}
{\caption{Heat maps for average log-damage of simulated storms in each ENSO phase. A different color scale is used for El Ni\~no to accentuate spatial patterns.}
\label{fig:simDam}}
\end{figure}

{To further decipher these effects, we generate {1,500} one-year storm forecasts for each ENSO phase from the {posterior predictive distribution}.} For storm paths, Figure \ref{fig:simPath} plots heat maps of simulated hit-rates in each location. Spatial patterns appear to be quite similar for each ENSO phase, with the two most susceptible locations being FL (simulated hit-rate 32.8\%, averaged over all ENSO phases), TX (21.9\%) and LA (16.1\%), whereas the least susceptible include ME (1.2\%), RI (1.8\%) and CT (2.3\%). This confirms the high spatial path correlations in Table \ref{tbl:post} and our observations in Section 2. Despite this similarity, Figure \ref{fig:simPath} provides some evidence for ENSO-spatial interactions: El Ni\~no and La Ni\~na storm simulations tend to travel further north (hitting the locations MA, RI, NY and CT 12.0\% and 9.0\% of the time, respectively) compared to Neutral storms (5.8\%), which validates our remarks for Figure \ref{fig:histpath}.\\

{For storm damage, Figure \ref{fig:simDam} plots heat maps of average log-damage for all simulated storms (multiple-location storms are included here, since the regional damage amounts $\{Y_s^{(i)}\}$ are available from simulations).} We see again that El Ni\~no storms incur notably less damage than both Neutral and La Ni\~na, with the latter two incurring similar damage on average. For spatial effects, a global pattern can be detected over all ENSO phases, with FL incurring higher damage in southern locations, and NY and CT incurring higher in the north. This suggests a \textit{shielding effect}: locations which are often the point of landfall (FL in the south, NY and CT in the north) receive the brunt of storm damage, while remaining locations are shielded from landfall and incur less damage when storm energy gradually dissipates over land (\citealp{Tul1994}). Figure \ref{fig:simDam} also provides visual evidence for ENSO-spatial interactions: El Ni\~no storms incur notably higher damage along the southern coast, most notably in GA (average log-damage = 19.0, 95\% HPD prediction interval [13.1, 24.7]) and FL (18.8, [13.9, 24.2]), whereas La Ni\~na storms incur higher damage along the northeast coast, most notably in NC (20.0, [14.8, 25.3]), VA (19.9, [14.8, 26.7]) and NY (19.9, [14.9, 25.1]). Since these northern locations generally have higher population densities, this supports our earlier hypothesis that higher-damaging storms (more prevalent in La Ni\~na) impact densely-populated locations more severely. Lastly, damage in Neutral storms experience much less spatial variation than La Ni\~na in Figure \ref{fig:simDam}, which is as observed in Table \ref{tbl:post}. This can be explained by the shielding effect: since La Ni\~na storms are more likely to make landfall in the higher density locations in the north, one expects these storms to incur higher damage for northern locations (and lower damage for southern locations) compared to Neutral storms.\\

Although the above observations are intuitive, they are not at all obvious from the exploratory data analysis in Section 2. By incorporating storm path data and using CAR models as spatial smoothers, our model can extract valuable insights on regional storm behaviour.

\subsection{Pricing regional premiums}
\begin{table}
\begin{floatrow}
\centering
\begin{tabular}{c c c c}
\toprule
{\textit{Location}} & {\textit{$VaR_{95\%}$}} & {\textit{$SV_{0.25}$}} & {\textit{$TVaR_{95\%}$}}\\
\toprule
\textit{ME} & 0.0 (0.0\%) & 0.2 (0.3\%) & 0.9 (0.4\%)\\
\textit{MA} & 0.2 (1.6\%) & 3.1 (5.7\%) & 8.3 (3.7\%)\\
\textit{RI} & 0.0 (0.0\%) & 0.9 (1.6\%) & 2.9 (1.3\%)\\
\textit{CT} & 0.0 (0.0\%) & 3.4 (6.3\%) & 8.4 (3.8\%)\\
\textit{NY} & 0.3 (2.3\%) & 5.3 (9.7\%) & 17.9 (8.0\%)\\
\textit{VA} & 0.0 (0.1\%) & 0.8 (1.4\%) & 2.9 (1.3\%)\\
\textit{NC} & 1.2 (9.5\%) & 6.3 (11.7\%) & 23.5 (10.6\%)\\
\textit{SC} & 0.6 (4.3\%) & 0.7 (1.2\%) & 4.0 (1.8\%)\\
\textit{GA} & 0.1 (0.5\%) & 1.7 (3.1\%) & 7.4 (3.3\%)\\
\textit{FL} & 4.9 (37.9\%) & 8.3 (15.3\%) & 43.5 (19.6\%)\\
\textit{AL} & 0.2 (1.3\%) & 1.3 (2.4\%) & 5.5 (2.5\%)\\
\textit{MS} & 0.6 (4.4\%) & 7.4 (13.8\%) & 26.6 (12.0\%)\\
\textit{LA} & 3.2 (24.4\%) & 13.2 (24.4\%) & 60.5 (27.2\%)\\
\textit{TX} & 1.7 (13.5\%) & 1.8 (3.3\%) & 10.0 (4.5\%)\\
\hline
\textbf{Total} & \textbf{13.0 (100\%)} & \textbf{54.2 (100\%)} & \textbf{222.5 (100\%)}\\
\toprule
\end{tabular}
{\caption{Annual regional premiums in bil. U.S. \$ (\% of total in brackets) in an El Ni\~no year using $SV_{0.25}$, $VaR_{95\%}$ and $TVaR_{95\%}$.}
\label{tbl:prem}
}
\end{floatrow}
\end{table}

In actuarial literature, a risk measure is a functional $\mathcal{H}$ mapping a loss random variable $X$ to the non-negative real numbers $\mathbb{R}_+$, where $\mathcal{H}$ is ``assumed in some way to encapsulate the risk associated with the loss" \citep{Har2006}. Three risk measures to illustrate the regional pricing process:
\bi
\item \textbf{Semi-variance ($SV_\eta$):} $\mathcal{H}(X) = \mu_X + \eta \sqrt{\sigma^2_{+}}, \; \eta \geq 0,$ where $\mu_X$ is the mean of X and $\sigma^2_+ = \mathbb{E}[\max(X-\mu_X,0)^2]$ is the {semi-variance} of $X$. $SV_\eta$ is motivated by the fact that an insurer's risk lies on the right-tail of $X$, so risk loadings should be calculated on only the positive portion of variance.
\item \textbf{Value-at-Risk ($VaR_\alpha$):} $\mathcal{H}(X) = F_X^{-1}(\alpha), \; 0 < \alpha < 1,$ where $F_X^{-1}(\alpha)$ is the $\alpha$-th quantile of $X$. When an insurer has $VaR_{\alpha}$ amount of capital available, it incurs a loss with at most probability $1-\alpha$.
\item \textbf{Tail-Value-at-Risk ($TVaR_q$):} $\mathcal{H}(X) = \mathbb{E}[X|X> F_X^{-1}(q)], \; 0 < q < 1.$ $TVaR_q$ ensures an insurer can cover the average loss given this loss exceeds its $q$-th percentile. 
\ei
The parameters $\eta$, $\alpha$ and $q$ quantify an insurer's risk-aversion level; larger values imply more risk-averse (and therefore higher) premiums, and vice versa. For illustration, we choose $\eta = 0.25$ and $\alpha = q = 95\%$. We further assume, for simplicity, that there are no deductibles, upper limits or coinsurance, and no discount interest rate for claims as well (these features can easily be added in by modifying MCMC predictions). P\&C (Property \& Casualty) insurers seldom use $VaR$ pricing in practice, since it, as a risk measure, is incoherent and fails to incorporate important risk information beyond quantiles. However, $VaR$ provides deeper insights on the other two pricing schemes, so we include it in our discussion.\\

\begin{figure}[!t]
\includegraphics[scale = 0.60]{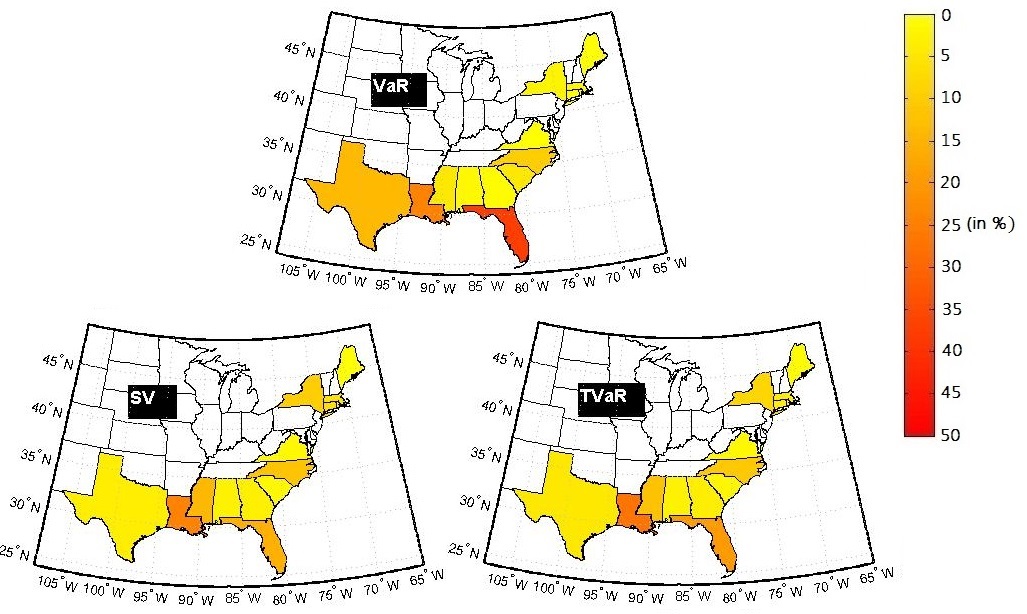}
\caption{\label{fig:prem}Colour map of premium allocation proportion for $SV_{0.25}$, $VaR_{95\%}$ and $TVaR_{95\%}$.}
\end{figure}

{Similar to the previous subsection, the MCMC samples from the full posterior distribution are used to estimate expectations and quantiles for $\mathcal{H}$.} Table \ref{tbl:prem} summarizes the $SV_{0.25}$, $VaR_{95\%}$ and $TVaR_{95\%}$ premiums for an El Ni\~no year. $VaR$ premiums (totalling \$13.0 bil.) are much lower than the other two measures in all locations, which is expected since $VaR$ pricing ignores heavy right-tailed losses exceeding quantiles. {On the other hand, by using conditional excess means to load premiums, $TVaR$ (totalling \$222.5 bil.) over-compensates for heavy-tailed losses, which results in unrealistic pricing schemes. $SV$ premiums (totalling \$54.2 bil.) appear to strike a good balance between the under-compensation of risk in $VaR$ with the over-compensation in $TVaR$.}\\

The spatial allocation of premiums is of primary interest to us, since it highlights the advantages of the proposed model over macro-level models. Figure \ref{fig:prem} plots the heat maps for the relative allocation {proportion} under each risk measure. In all three maps, FL and LA are charged a majority of total premiums, which is expected since these locations have the highest simulated storm hit-rates (Figure \ref{fig:simPath}). $SV$ and $TVaR$, however, provide a noticeable spatial smoothing on $VaR$, in that locations with high allocation {proportion} under $VaR$ tend to have slightly lower {proportion} under $SV$ and $TVaR$. There are two reasons for this smoothing. First, although all risk measures incorporate both path effects and damage effects, $VaR$ prioritizes path effects by pricing on only quantiles. For example, since FL incurs lower damage-per-storm than NC but has higher hit-rates, it is charged a higher allocation of premiums by $VaR$ than by $SV$ or $TVaR$. Another reason is that the storm severity effect $\hat{\sigma}^2_{\zeta} = 1.66$, which affects all locations along a path, produces similar right-tail loss behaviour for LOS neighbours. For instance, the premium allocation for FL is smoothed much more than for LA, since the former has more LOS neighbours than the latter.


\section{Discussion}
After a period of intense hurricane activity, most notably Hurricane Katarina in 2005, attention has shifted to understanding the nature and economic impact of these storms. The compound Poisson-lognormal process \eqref{eq:CPL} was proposed for aggregate claims, but cannot serve as a micro-level model since it does not account for underlying storm paths. From an application standpoint, we develop methodology that allows for localized prediction of HTS damage and thus regional pricing of insurance products.  A key innovation in the model specification is the use of LOS connectivity rather than physical connectivity, for that not only allows for regional path information to be shared between more locations, but also provides more path predictions that are physically possible. The proposed model could be useful for pricing regional insurance premiums, and possibly for guiding policies on disaster prevention as well. \\

For posterior exploration of each submodel, a hybrid Metropolis-within-Gibbs sampler is used which exploits direct conditional sampling whenever possible. Simulation studies show adequate MCMC mixing, accurate parameter inference, and good regional predictions using aggregate damage data. When fit to historical data, the proposed model reveals several interesting insights, some new and others confirming previous findings. For storm count, our analysis shows that colder SST anomalies are linked with higher storm counts (as noted in \citealp{PL1998} and \citealp{Kat2002}), and provides evidence for varying seasonal patterns by ENSO (as mentioned in \citealp{LZ2012}). For storm path, a common spatial pattern is observed over all ENSO phases, with a noticeable higher tendency for El Ni\~no and La Ni\~na storms to hit northern locations. For storm damage, the fitted model supports the association between colder anomalies and higher storm damage, and also reveals a changing spatial pattern for damage, with El Ni\~no storms incurring higher damage in southern locations and La Ni\~na incurring higher damage in the north.\\

For predictions, we assumed for simplicity that the ENSO phase for the upcoming year is known. This is not the case in reality: according to IRI \footnote{http://iri.columbia.edu/our-expertise/climate/forecasts/enso/current/}, one-year SST forecasts from 25 common forecasts may vary by as much as $1.5 \,^{\circ}\mathrm{C}$! A natural extension of the proposed model is to quantify ENSO phase uncertainty through an appropriate stochastic model. Another future direction is to speed up computation for posterior exploration, since generating 20,000 MCMC samples from historical data takes over 4 hours on a quad-core Ivy Bridge 2.7 GHz processor desktop (with code written in MATLAB version 8.1). After a profile inspection, the bottleneck step is computing the denominator of \eqref{eq:ALjointnormed}, which requires summing over all LOS connected paths. Several approximations have been proposed (e.g., \citealp{Hea2000} and \citealp{Hue2011}), and it will be interesting to see whether these computational speed-ups require a substantial tradeoff in sampling accuracy.\\

Our proposed model \eqref{eq:Xtagg} can be easily extended to model not only economic loss from natural disasters with an underlying physical path, such as tornados, blizzards and wildfire, but also in certain weather and climate models. One simply has to identify potential covariates (such as ENSO and seasonality) and model spatial interactions for a suitable subset of these covariates (determined through prior knowledge or exploratory analysis). Lognormal losses may not be appropriate for some applications, and other thinner- or thicker-tailed distributions may be more suitable instead, e.g., gamma or generalized extreme value (GEV) distributions. The framework in \eqref{eq:Xtagg}, however, provides a flexible micro-level model which not only provides regional predictions by allowing information to be borrowed between spatial neighbours, but also reveals useful insights on spatial patterns which would otherwise be lost from only a macro-level analysis.


\footnotesize{\bibliography{references}}

\normalsize
\begin{appendices}
\numberwithin{equation}{section}
\counterwithin{figure}{section} 

\section{Map and legend}

\begin{figure}[H]
\centering
\includegraphics[scale = 0.45]{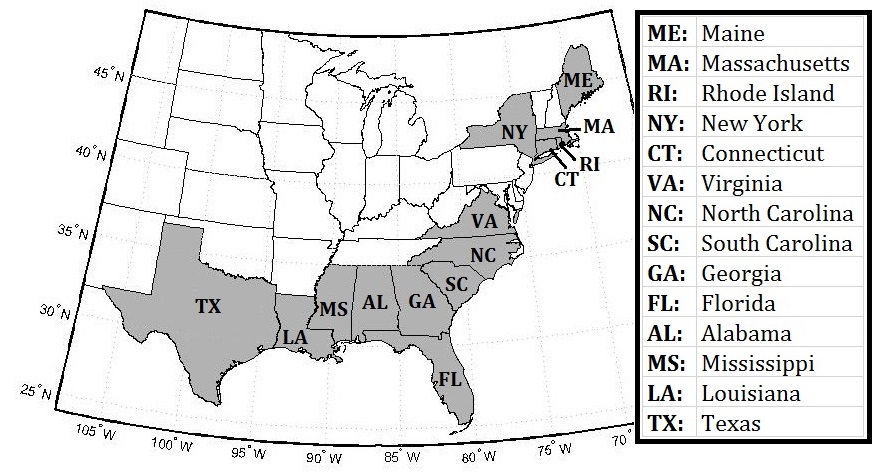}
\caption{Map and legend of U.S. states.}
\label{fig:legend}
\end{figure}

\section{Derivation of full conditionals}

Here, we derive the full conditional distributions of $\bm{\Sigma}_\gamma$, $\bm{\Sigma}_\xi$, $\sigma^2$ and $\sigma^2_\zeta$ in the storm path and damage samplers. For the path sampler, let $\mathcal{P} = \{\boldsymbol{\delta}^{(i)}: i = 1, \cdots, n\}$ be the storm path data for $n$ observed storms. Using the likelihood \eqref{eq:MCARjoint2} along with the prior specification in Table \ref{table:hiermodel}, the joint posterior of the path submodel parameters $\Theta^P$ becomes
{\small
\begin{align}
\begin{split}
[\Theta^P|\mathcal{P}] \propto & \left[\prod_{i = 1}^n p(\boldsymbol{\delta}^{(i)}|\boldsymbol{\gamma}_0, \boldsymbol{\gamma}, \phi) \right] \cdot \left[|\bm{\Sigma}_\gamma|^{-\textrm{rank}(\bm{D}_w - \bm{W})/2}\exp\left\{ -\frac{1}{2} \boldsymbol{\gamma}^T[(\bm{D}_w - \bm{W}) \otimes \bm{\Sigma}_{\gamma}^{-1}]\boldsymbol{\gamma}\right\}\right]\\
& \cdot \frac{1}{\phi} \cdot \left[|\bm{\Sigma}_\gamma|^{-4} \exp\left\{ -\frac{1}{2} \textrm{tr}(\bm{\Sigma}^{-1}_\gamma) \right\}\right],
\end{split}
\label{eq:pathlkhd}
\end{align}
}
where $\textrm{tr}(\cdot)$ is the trace function. Let $\boldsymbol{\gamma'} = (\boldsymbol{\gamma}_{1,\cdot}^T, \boldsymbol{\gamma}_{2,\cdot}^T, \boldsymbol{\gamma}_{3,\cdot}^T)^T$ be a rearranged vector for $\boldsymbol{\gamma}$, and, with a slight abuse of notation, let $\ast = \Theta^P \setminus \theta$, where $\theta$ is the current parameter considered. From \eqref{eq:pathlkhd}, the full conditional distribution of $\bm{\Sigma}_\gamma$ becomes
\begin{align}
\begin{split}
[\bm{\Sigma}_\gamma|\mathcal{P}, \ast]  
& \sim IW \left( 3 + \textrm{rank}(\bm{D}_w - \bm{W}), \boldsymbol{\gamma'}^T (\bm{D}_w-\bm{W}) \boldsymbol{\gamma'} + \bm{I}\right).
\end{split}
\label{eq:cond1}
\end{align}

Likewise, for the damage sampler, let $\boldsymbol{\xi}_{\cdot,s} \equiv (\xi_{1,s}, \xi_{2,s}, \xi_{3,s})^T$ for each location $s = 1, \cdots, S$, and let $\mathcal{D} = \{\boldsymbol{\delta}^{(i)}, Y^{(i)}_s: i = 1, \cdots, n; s = 1, \cdots, S\}$ be the set of path data and damage incurred in each location. Using \eqref{eq:MCARjoint2} and \eqref{eq:myLNmodel} along with the prior specification in  Table \ref{table:hiermodel}, the joint posterior for the damage submodel parameters $\Theta^D$ and storm severity parameters $\{\zeta^{(i)}\}_{i = 1}^n$ becomes
{\small 
\begin{align}
\begin{split}
[\Theta^D \cup \{\zeta^{(i)}\}|\mathcal{D}] & \propto \left[\prod_{i = 1}^n \prod_{s:\delta_s^{(i)} = 1}\frac{1}{Y^{(i)}_s \sigma} \;\exp\left\{-\frac{(\ln Y^{(i)}_s - \xi_{k(t_i)}^0 - \xi_{k(t_i),s} - \zeta^{(i)})^2}{2 \sigma^2}\right\}\right]\\
& \cdot \left[|\bm{\Sigma}_\xi|^{-\textrm{rank}(\bm{D}_w - \bm{W})/2}\exp \left\{ -\frac{1}{2} \boldsymbol{\xi}^T[(\bm{D}_w - \bm{W}) \otimes \bm{\Sigma}_{\xi}^{-1}]\boldsymbol{\xi} \right\} \right] \cdot \left[ \prod_{i = 1}^n \frac{1}{\sigma_\zeta} \exp\left\{-\frac{{\zeta^{(i)}}^2}{2\sigma^2_\zeta}\right\} \right]\\
& \cdot \left[ (\sigma^2)^{-0.01-1} \exp\left\{ -\frac{0.01}{\sigma^2} \right\} \right] \cdot \left[ |\bm{\Sigma}_\xi|^{-4} \exp\left\{ -\frac{1}{2} \textrm{tr}(\bm{\Sigma}^{-1}_\xi) \right\} \cdot (\sigma^2_\zeta)^{-0.01-1} \exp \left\{ -\frac{0.01}{\sigma^2_\zeta} \right\} \right].
\end{split}
\label{eq:damagejoint}
\end{align}
}
Let $\boldsymbol{\xi}' = (\boldsymbol{\xi}_{1,\cdot}^T, \boldsymbol{\xi}_{2,\cdot}^T, \boldsymbol{\xi}_{3,\cdot}^T)^T$ be a rearranged vector for $\boldsymbol{\xi}$. From \eqref{eq:damagejoint}, the full conditional distribution of $\bm{\Sigma}_\xi$ becomes
\begin{align}
\begin{split}
[\bm{\Sigma}_\xi|\mathcal{D}, \ast] & \sim IW \left( 4 + \textrm{rank}(\bm{D}_w - \bm{W}), \boldsymbol{\xi}'^T (\bm{D}_w-\bm{W}) \boldsymbol{\xi}'+ \bm{I} \right),
\end{split}
\label{eq:cond2}
\end{align}
the full conditional distribution of $\sigma^2$ becomes
\begin{align}
\begin{split}
[\sigma^2|\mathcal{D}, \ast] 
& \sim {IG} \left( \left\{\sum_{i = 1}^n \sum_{s = 1}^S \bm{1}_{\{\delta_s^{(i)} = 1\}}\right\}/2 + 0.01, \left\{\sum_{i = 1}^n \sum_{s:\delta_s^{(i)} = 1} (\ln Y^{(i)}_s - \xi_{k(t_i)} - \xi_{k(t_i),s} - \zeta^{(i)})^2\right\}/{2} + 0.01\right),
\end{split}
\label{eq:cond4}
\end{align}
the full conditional distribution of $\sigma^2_\zeta$ becomes
\begin{align}
\begin{split}
[\sigma^2_\zeta|\mathcal{D}, \ast] 
& \sim {IG} \left( \frac{n}{2} + 0.01, \frac{\sum_{i = 1}^n {\zeta^{(i)}}^2}{2} + 0.01\right),
\end{split}
\label{eq:cond3}
\end{align}
and the full conditional distributions of storm severities parameters $\{\zeta^{(i)}\}_{i=1}^n$ become
\begin{align}
\begin{split}
[\zeta^{(i)}|\mathcal{D}, \ast]
& \sim N\left( \frac{\sigma^2_\zeta}{\sigma_\zeta^2 (\boldsymbol{\delta}^{{(i)}^T} \boldsymbol{1}) + \sigma^2}\sum_{s:\delta_s^{(i)} = 1}\left( \ln Y^{(i)}_s - \xi_{k(t_i)} - \xi_{k(t_i),s} \right), \frac{\sigma^2 \sigma^2_\zeta}{\sigma_\zeta^2 (\boldsymbol{\delta}^{{(i)}^T} \boldsymbol{1}) + \sigma^2} \right).
\end{split}
\label{eq:cond5}
\end{align}

{\section{Conditional sampling of $\{Y_s^{(i)}\}$ given $Y^{(i)}$}
Here, we discuss an approximate algorithm for sampling the latent regional damage of a multiple-location storm, given its total incurred damage over all locations. For notational convenience, we consider only one multiple-location storm occuring in ENSO phase $k$, and drop the superscript $i$. With a slight abuse of notation, let $1, 2, \cdots, M$ be new indices for the set of affected locations $\{s:\delta_s=1\}$, let $Y$ be the total incurred damage, and let $Y_{m+} = \sum_{l = m+1}^M Y_l$ be the sum of regional damage with new index exceeding $m$. Following \cite{Mea2007}, we approximate the density of $Y_{m+}$ with a lognormal density having parameters $\mu_{m+}$ and $\sigma^2_{m+}$ (denoted as $f_{LN}(z;\mu_{m+}, \sigma^2_{m+})$), with $\mu_{m+}$ and $\sigma^2_{m+}$ obtained by matching the mgf of $Y_{m+}$ with the mgf of the estimated lognormal distribution at the points $-0.001$ and $-0.005$ (suggested in \citealp{Mea2007}). {The desired distribution can be written as
\begin{equation}
[Y_1, \cdots, Y_{M}|Y, \boldsymbol{\Theta}^D] = [Y_1 | Y, \boldsymbol{\Theta}^D] \cdot [Y_2| Y_1, Y, \boldsymbol{\Theta}^D] \cdots [Y_{M-1}| Y_1, \cdots, Y_{M-2}, Y, \boldsymbol{\Theta}^D] \cdot \bm{1}{\left\{ \sum_{l =1}^{M-1} Y_l \leq Y \right\}},
\label{eq:condpart}
\end{equation}
with each term on the right-hand side of \eqref{eq:condpart} approximated by:
\begin{align}
\begin{split}
[Y_1 = z| Y, \boldsymbol{\Theta}^D] & \propto f_{LN}(z; \xi^0_k + \xi_{k,1} + \zeta, \sigma^2) \cdot f_{LN}(Y-z; \mu_{1+},\sigma^2_{1+})\\
[Y_2 = z| Y_1, Y, \boldsymbol{\Theta}^D] &\propto f_{LN}(z;\xi^0_k + \xi_{k,2} + \zeta, \sigma^2) \cdot f_{LN}(Y - Y_1 - z; \mu_{2+},\sigma^2_{2+})\\
& \vdots \\
[Y_{M-1} =z | Y_1, \cdots, Y_{M-2}, Y, \boldsymbol{\Theta}^D] &\propto f_{LN}(z; \xi^0_k + \xi_{k,M-1} + \zeta, \sigma^2) \; \cdot\\
& \quad \quad  f_{LN}\left(Y - \sum_{l =1}^{M-2} Y_l - z; \mu_{(M-1)+},\sigma^2_{(M-1)+} \right).
\end{split}
\label{eq:approx}
\end{align}
\\
A Metropolis sampler can then be implemented which proposes $Y_1, \cdots, Y_{M-1}$ in a single block (rejecting whenever $Y_l < 0$ for some $l = 1, \cdots, M-1$ or $\sum_{l=1}^{M-1} Y_l > Y$), and accepts proposals with probabilities computed from \eqref{eq:condpart} and \eqref{eq:approx}. For better mixing of this sampler, we first scale the sample space of $[Y_1, \cdots, Y_M|Y, \boldsymbol{\Theta}^D]$ to the unit simplex, and use an independent $U[0,1]^{M-1}$ random walk (with step-size tuned using the algorithm in \citealp{Gra2011}) for proposals. Note that this algorithm may indeed experience poor mixing when $M$, the number of locations hit by the same storm, is large. For the HTS data, however, neither $M$ nor the number of spatial locations considered are large ($M\leq 4$ for all storms), so this is not a concern for our application.
}
}
\end{appendices}


\end{document}